\documentclass[fleqn,usenatbib]{mnras}

\usepackage{newtxtext,newtxmath}
\usepackage[T1]{fontenc}

\usepackage{graphicx}	% Including figure files
\usepackage{amsmath}	% Advanced maths commands
\usepackage[normalem]{ulem}
\usepackage{bm}

\usepackage{color}

\title[$f\left(R\right)$ gravity versus the Hubble tension]{Metric $f(R)$ gravity with dynamical dark energy as a 
scenario for the Hubble tension}

\author[G. Montani \& M. De Angelis \& F. Bombacigno \& N. Carlevaro]{
Giovanni Montani,$^{1,2}$\thanks{E-mail: giovanni.montani@enea.it}
Mariaveronica De Angelis,$^{3}$\thanks{E-mail: mdeangelis1@sheffield.ac.uk}
Flavio Bombacigno,$^{4}$\thanks{E-mail: flavio2.bombacigno@uv.es}
Nakia Carlevaro,$^{1}$\thanks{E-mail: nakia.carlevaro@enea.it}
\\
$^{1}$ENEA, Fusion and Nuclear Safety Department, C.R. Frascati, Via E. Fermi 45,  Frascati, I-00044 Rome, Italy\\
$^{1}$Physics Department, ``Sapienza" University of Rome, P.le Aldo Moro 5, I-00185 Rome, Italy\\
$^{3}$School of Mathematics and Statistics, University of Sheffield, Hounsfield Road, Sheffield S3 7RH, United Kingdom\\
$^{4}$Departament de F\'{i}sica Teòrica and IFIC, Universitat de València, Carrer del Doctor Moliner 50, 46100, Burjassot, Spain\\
}

\date{Accepted XXX. Received YYY; in original form ZZZ}

\pubyear{2022}

\begin{document}
\label{firstpage}
\pagerange{\pageref{firstpage}--\pageref{lastpage}}
\maketitle

% It should be a single paragraph not more than 250 words (200 words for Letters).
% No references should appear in the abstract.
\begin{abstract}

We introduce a theoretical framework to interpret the Hubble tension, based on the combination of a metric $f(R)$ gravity with a dynamical dark energy contribution. The modified gravity provides the non-minimally coupled scalar field responsible for the proper scaling of the Hubble constant, in order to accommodate for the local SNIa pantheon+ data and Planck measurements. The dynamical dark energy source, which exhibits a phantom divide line separating the low red-shift quintessence regime ($-1<w<-1/3$) from the phantom contribution ($w<-1$) in the early Universe, guarantees the absence of tachyonic instabilities at low red-shift. The resulting $H_0(z)$ profile rapidly approaches the Planck value, with a plateau behaviour for $z\gtrsim 5$. In this scenario, the Hubble tension emerges as a low red-shift effect, which can be in principle tested by comparing SNIa predictions with far sources, like QUASARS and Gamma Ray Bursts.
\end{abstract}


%%%%%%%%%%%%%%%%%%%%%%%%%%%%%%%%%%%%%%%%%%%%%%%%%%

%%%%%%%%%%%%%%%%% BODY OF PAPER %%%%%%%%%%%%%%%%%%
\section{Introduction}

Modern Cosmology (\cite{bib:weinberg-2008}) suffers from a significant number of unexplained features, among which the problem of the late Universe acceleration (dark energy) stands for its relevance on the incoming observational tasks (see, for instance \cite{Amendola:2016saw,Huterer:2017buf,Mortonson:2013zfa,Kase:2018aps,Frusciante:2019xia}). 

Many different scenarios have been conjectured on the nature of the present acceleration and they could be distinguished into two main classes. On one hand, we have models requiring the presence of exotic sources, leading to matter with negative pressure mimicking vacuum energy density contribution; on the other hand, we can consider modified gravity effects able to reproduce the same phenomenology via non-Einsteinian dynamics (\cite{CLIFTON20121}). 

Recently, this puzzling picture has been enriched by the so-called $4.9 \sigma$ \lq \lq Hubble tension" (\cite{DiValentino:2021izs}), consisting of the non-concealable discrepancy between the value of the Hubble constant $H_0$ (henceforth measured in $\text{km\,s}^{-1} \text{Mpc }^{-1} $) by the Planck satellite ($67.4 \pm 0.5$) (\cite{Planck:2018vyg}) and by the nearby standard candles like Supernovae Ia, hereafter SNe Ia ($73.3 \pm 1.1$) (\cite{Brout:2022vxf}). This issue suggests that the Hubble constant measurement could be affected by the red-shift values at which data are taken.  

Furthermore, the analysis provided in \cite{Dainotti:2021pqg} (see also \cite{galaxies10010024,kazantzidis,Krishnan:2020obg}) outlines the existence of a dependence of the type $H_0(z) \propto (1 + z)^{-\alpha}$ where $\alpha \simeq 10^{-2}$ from a red-shift binned analysis of the SNIa distribution in the Pantheon sample (\cite{Pan-STARRS1:2017jku}). This result has been obtained within $2 \sigma$ of statistical significance and it suggests that the Hubble tension can be intended as an effective monotonic decreasing of the $H_0$ value from $z=0$ up to $z\simeq 1100$.

This behaviour for the Pantheon sample could be explained by a red-shift evolution of the SNe Ia as astrophysical sources, and, hence, as standard candles. However, this perspective is not entirely shared by the community investigating these sources (see for instance the Pantheon$+$ analysis presented in \cite{Brout:2022vxf}), and it seems legit to interpret the $H_0(z)$ profile of the Pantheon SNIa sample as the Hubble tension in its-self, by introducing a new background physics. In this respect, in \cite{Dainotti:2021pqg} it was argued that the profile $H_0(z)$ is the result of the Einstein constant rescaling via the non-minimally coupled scalar field of metric $f(R)$ gravity in the Jordan frame (for other approaches following similar paradigms, see \cite{Odintsov:2020qzd,Banerjee:2022ynv,Nojiri:2022ski}). 

However, a Hu-Sawicki model was tested in \cite{galaxies10010024}, and the associated luminosity distance turned out to not account for the desired effect. This is essentially due to the typical behaviour of the modified gravity theories providing a Universe acceleration: they reproduce a $\Lambda$CDM model (\cite{bib:weinberg-2008}) only at large red-shift, while small deviations are today available.
This feature is quite in contradiction with the aim of dealing with the Hubble tension, thought as a $H_0(z)$ profile, for which the deviation from the standard $\Lambda$CDM model must take place just at large values of the red-shift. This has been successfully implemented in \cite{schiavone_mnras}, where a metric $f(R)$ gravity model able to reproduce the late Universe acceleration and the $H_0(z)$ profile in \cite{Dainotti:2021pqg} was constructed. A smooth decreasing behaviour of $H_0(z)$ is indeed obtained, matching with the detected values of the SNIa and the Cosmic Microwave Background (CMB) data. As originally conjectured, it is just the non-minimally coupled scalar field of the scalar-tensor representation to be responsible for the Hubble constant scaling, making it apparently depending on the red-shift of the sources used for its determination. 

Here, we want to address the problem of the Hubble tension by searching for a physical effect able to rescale the Universe expansion rate but without requiring a priori the specific dependence on the red-shift $z$ introduced in \cite{schiavone_mnras}. %Within the context of metric $f(R)$ gravity, we study the red-shift evolution of the dynamical system for the following unknown: 
Within the context of metric $f(R)$ gravity, we examine the evolutionary behaviour of the red-shift for a dynamical system involving three variables that are not yet determined: the Hubble function, the non-minimally coupled scalar field, and its potential term. %In particular, in order to ensure the physical (non-tachyonic) character of the model, we introduce, in addition to the standard matter energy density contribution, a dynamical dark energy component exhibiting a state parameter running with the red-shift
To prevent the model from having tachyonic properties, we incorporate a dynamic dark energy component alongside the conventional matter energy density contribution. This dark energy component displays a varying state parameter that correlates with the red-shift (see also \cite{Heisenberg:2022gqk,Heisenberg:2022lob,PhysRevD.103.L081305,Zhao:2017cud,PhysRevD.98.083501,PhysRevD.104.023510,Adil:2023exv,DiValentino:2022eot,Akarsu:2023mfb,Krishnan:2020vaf,Malekjani:2023dky}). The model we construct is able to provide the desired rescaling of the Hubble constant. The function $H_0(z)$ rapidly decreases from the SNIa value to the CMB one, which is already reached at $z\simeq 5$ (even at $z\gtrsim 2$ if error bars are considered). 

The necessary modified gravity theory deviates from General Relativity only at low red-shifts, and Einstein-Hilbert action in the presence of a cosmological constant (up to a rescaling of the gravitational constant) is promptly recovered. The dynamical dark energy is driven by two parameters, ruling the transition from quintessence regime to a phantom energy phase ($w<-1$), with the phantom divide line (PDL) (\cite{Zhou:2021xov,Escamilla:2023shf,Escamilla:2021uoj,Hojjati:2009ab,PhysRevD.74.083521,DiValentino:2020naf}) occurring around $z\simeq 0.83$. The parameters are determined within certain ranges, depending on the error bars of the $H_0$ measurements at low and high red-shifts.

The most important phenomenological feature of our model is that it discriminates between the SNe Ia as testers, through whose sample $H_0(z)$ rapidly decays (also according to the analysis in \cite{Dainotti:2021pqg,galaxies10010024}) and higher red-shift sources, like QUASARS (\cite{Colgain:2022nlb,Dainotti:2022rfz,Bargiacchi:2023jse}) and Gamma Ray Bursts (\cite{Dainotti:2022wli,10.1093/pasj/psac057,Cao:2022wlg,Cao:2022yvi,10.1093/mnras/stac2752}), for which the value of $H_0$ has to coincide with the CMB measured one. This fact offers an interesting validation perspective of the proposed theoretical framework. 

The paper is organized as follows. In Sec.~\ref{sec: sec. 2} we introduce the metric $f(R)$ theory of gravity and we apply it to the late universe. This implementation results in a generalized Friedmann equation that encompasses the presence of both baryonic and non-baryonic matter, along with dynamic dark energy fluid. The aim of Sec.~\ref{sec: sec. 3} is to find a dynamical framework which could be understood as a $\Lambda$CDM model, wherein the Hubble constant $H_0$ exhibits a red-shift-dependent variation. This feature arises as a consequence of the evolution of the non-minimally coupled scalar field $\phi(z)$. Moreover, we express the equation of state for the dynamical dark energy. In Sec.~\ref{sec: sec. 4} we show how the presence of a dynamical dark energy source is a viable tool to ensure the absence of tachyonic instabilities which could otherwise impact the modified gravity model under consideration at low red-shifts. Finally, it is discussed the $f(R)$ profile which manifests small deviations from General Relativity only at low-red-shift. In Sec.~\ref{sec: sec. 6} conclusions are drawn.
Spacetime signature is chosen mostly plus $(-,+,+,+)$, and we adopt units for $c=1$ with $\chi$ denoting the Einstein constant.

\section{Cosmological formulation for modified gravity}\label{sec: sec. 2}

Here, we consider a metric $f(R)$-model as described in the Jordan frame (\cite{Sotiriou-Faraoni:2010,NOJIRI201159}), in which the non-Einsteinian features are summarized by the presence of a non-minimally coupled scalar field $\phi$. The action of the theory takes the form
\begin{equation}
	S = \frac{1}{2\chi}\int d^4x 
	\sqrt{-g}\left( \phi R - V + 2\chi \mathcal{L}_m\right),
	\label{eq: action}
\end{equation}
where $g$ and $R$ are the determinant and Ricci scalar associated with the metric tensor $g_{\mu\nu}$ respectively, and $\mathcal{L}_m$ denotes the matter Lagrangian density. Here, the potential $V(\phi)$ is fixed by the specific functional form $f(R)$, according to the following relation
\begin{equation}
	V(\phi ) \equiv \phi R(\phi )- f(R(\phi )) \, , \quad R(\phi ) = 
	\left( \frac{df}{dR}\right) ^{-1},
	\label{rt2}
\end{equation}
being $df/dR$ an invertible function. It is worth noting that $g_{\mu\nu}$ and $\phi$ are independent degrees of freedom with respect to which the action (\ref{eq: action}) has to be varied, considering also the matter variables. In particular, the scalar field is a massive propagating mode that could be detected in the polarisation modes of a gravitational wave, see \cite{PhysRevD.100.084014}.

%\subsection{Cosmological implementation}

We now implement the modified gravity picture depicted above to the late Universe dynamics in which we neglect both the spatial curvature and the radiation contribution (\cite{bib:weinberg-2008}). Thus it is well described by a line element of the Robertson-Walker form 
\begin{equation}
	ds^2 = -dt^2 + a(t)^2\left( dx^2 + dy^2 + dz^2\right),
	\label{rt3}
\end{equation}
where we adopted Cartesian coordinates and the dynamics of the cosmic scale factor $a$ is expressed via the synchronous time $t$.
The generalized Friedmann equation is then obtained by considering the $tt$-component of the metric field equation derived from \eqref{eq: action}, resulting in:
\begin{equation}
    H(t)^2 = \frac{1}{3\phi}\biggl(
	\chi \rho - 3H\dot{\phi} + \frac{V}{2}\biggl)\,,
	\label{eq: gen Friedmann equation}
\end{equation}
where $H(t)\equiv \dot{a}/a$ is the Hubble parameter (here and in the following the dot denotes time differentiation) and $\rho$ represents the total energy density associated with the perfect fluid stress-energy tensor for $\mathcal{L}_m$. We set $\rho$ to accommodate for the non-relativistic component $\rho_m(t)=\rho_{m0}(a_0/a)^{-3}$ describing baryonic and non-baryonic matter (where $\rho_{m0}$ denotes the present day value of the mass-energy density and $a_0$ the corresponding value of the scale factor), as well as for a dynamical dark energy fluid characterized by the equation of state $P_\Lambda=w(t)\rho_{\Lambda}$, with $w(t)$ to be specified (see the discussion below). Eventually, by varying \eqref{eq: action} w.r.t. $\phi$, we get the relation
\begin{equation}
	\frac{dV}{d\phi} = 6\dot{H} + 12 H^2
	\, .
	\label{eq: scalar field equation}
\end{equation}
The dynamical system is described by \eqref{eq: gen Friedmann equation} and \eqref{eq: scalar field equation}, which together with the continuity equation for the matter
\begin{equation}
    \dot \rho+3H(\rho+P)=0,
    \label{eq: continuity}
\end{equation}
completely determines the evolution of the degrees of freedom $a(t),\;\phi (t)$ and $\rho(t)$ (once $w(t)$ is assigned).

\section{Construction of the model}\label{sec: sec. 3}

Within the considered model above, we now search for a solution able to reproduce the current Universe expansion and the apparent variation of the Hubble constant related to the Hubble tension. In other words, we are searching for a dynamical paradigm that can be interpreted as a $\Lambda$CDM model in which Hubble constant $H_0$ acquires an effective dependence on the red-shift. We will get this feature as the effect of the scalar field evolution 
$\phi = \phi (z)$. 

It is convenient to rearrange all the quantities in terms of the red-shift-like variable $x\equiv \ln (1+z)$. Here $z$ is defined from 
\begin{equation}
a_0/a=1+z,
\end{equation}
where $a_0$ represents the present-day scale factor. Using the convention $a_{0}=1$, differentiation in $t$ can be expressed as
\begin{equation}
	\frac{d}{dt}=-H(x)\frac{d}{dx}.
	\label{rt8}
\end{equation}

Having in mind a scenario that is very close to a $\Lambda\text{CDM}$ model, we impose that the potential cancels out (up to a rescaling of the gravitational coupling) the modified gravity contribution led by the scalar field, i.e.
\begin{equation}
	V(\phi ) =6H\dot{\phi}, 
	\label{rt9}
\end{equation}
and we choose, for the dynamical dark energy component, the equation of state
\begin{equation}
    P_\Lambda(x)=-\biggl(1-\frac{w_0+2w_1 x}{3}\biggl)\rho_\Lambda(x).
\end{equation}
Here, the parameters $w_0$ and $w_1$ account for deviations from the standard cosmological constant scenario, which is recovered for $w_0=w_1=0$. At low red-shifts, where $x\simeq z$, this equation of state reproduces the so-called $w_0w_a$CDM model (\cite{Pan-STARRS1:2017jku}). In particular, as shown in Fig.~\ref{rho e omega} our expression exactly coincides with a $w_0w_a$ model for $z\lesssim 0.2$. Such a dynamical dark energy contribution, once $w_0$ and 
$w_1$ are properly determined (see discussion in Sec.~\ref{sec: sec. 4}), is necessary to guarantee a consistent $f(R)$ gravity devoid of tachyonic degrees of freedom. Taking into consideration \eqref{eq: continuity}, it follows that \eqref{eq: gen Friedmann equation} can be rewritten as
\begin{equation}
	H^2(x) = \frac{\bar{H}_0^2}{\phi(x)} 
	\left( \Omega_{m0}e^{3x} + \Omega _{\Lambda0}e^{x(w_0+w_1 x)}\right), 
	\label{rt10}
\end{equation}
where $\bar{H}_0$ is taken as the Hubble constant measured by the SNe Ia (i.e. $\bar{H}_0=H_0(z=0)$). Moreover, $\Omega_{m0}$ and $\Omega _{\Lambda0}$ denote the present-day values of the matter and vacuum energy density critical parameters, respectively. Specifically, we use the following definitions: $\Omega_{m0}=\rho_{m0}/\rho_{c0}$ and $\Omega_{\Lambda0}=\rho_{\Lambda0}/\rho_{c0}=1-\Omega_{m0}$ (flat Universe), where $\rho_{c0}$ is the critical energy density of the Universe today taken as $\rho_{c0}=3\bar{H}_{0}^{2}/\chi$. Similarly to the $\Lambda\text{CDM}$ model, from \eqref{rt10} we can now define the following time-dependent effective Hubble constant as
\begin{equation}\label{heffphi}
H_{0,\text{eff}}(x)=\frac{\bar{H}_{0}}{\sqrt{\phi(x)}}\,,
\end{equation}
where we naturally fix $\phi(0)=1$ to obtain $H_{0,\text{eff}}(0)=\bar{H}_0$.

Finally, expressing \eqref{eq: scalar field equation} and (\ref{rt9}) via $x$, we now get
\begin{align}
	\frac{dV(x)}{dx} &=\biggl( 12H^2-6H\frac{dH}{dx}\biggl)\frac{d\phi}{dx}
	\, ,
	\label{rt12}\\
    V(x) &= - 6H^2 \frac{d\phi}{dx}.
	\label{rt11}
\end{align}

The three equations (\ref{rt10}), (\ref{rt12}) and (\ref{rt11}) can be solved for the three unknowns $H(x)$, $\phi(x)$ and $V(x)$. Furthermore, combining the last two functions, we can also infer the profile $V(\phi )$ and, hence, the nature of the considered $f(R)$ gravity.

\section{Toward the solution for the Hubble tension}\label{sec: sec. 4}
We start our analysis by observing that the ratio of (\ref{rt12}) and (\ref{rt11}) results in
\begin{equation}
	\frac{d \ln V}{d x} = - 2 + \frac{d\ln H}{d x}
	\, , 
	\label{rt13}
\end{equation}
which admits the solution
\begin{equation}
	V(x) = 
	\frac{H(x)}{\lambda e^{2x}}, 
	\label{rt14}
\end{equation}
where we fixed the integration constant as $V(0)\equiv H(0)/\lambda$, with $\lambda$ a negative constant with dimension of length. Now, comparing the equation above with \eqref{rt11} and taking into account \eqref{rt10}, we get 
\begin{equation}
	 \frac{d \phi}{dx}=-\frac{1}{6\lambda \bar{H}_{0}e^{2x}}\sqrt{\frac{\phi(x)}{\Omega_{m0}e^{3x}+(1-\Omega_{m0})e^{x(w_0+w_1 x)}}}.
	\label{rt15}
\end{equation}
%where we must retain the positive root according to the expanding behaviour of the present Universe. 

%\mary{in che senso? la distanza di luminosità si calcola per estrapolarsi il valore di H0 alla CMB perche si conoscono gli altri due parametri che sono l'angolo e r. Se la distanza in luminosita cambia vuol dire che il valore di H0 alla CMB cambia per questo modello che abbiamo qui ma la differenza di H0 tra SN e CMB resta. Non ho capito perchè hai scritto "that could remove the apparent variation of the HUbble constant". cioè il paradigma è soddisfatto ma non ho capito perchè dici che elimina la variazione, forse bisogna rifrasarlo }

The behaviour of $\phi(x)$\footnote{It is worth stressing that, from an observational point of view, the presence of the scalar field $\phi$ emerges via an integrated effect in the photon path. In fact, it enters the luminosity distance $d_l$ that, calculated in a generic instant $\bar{x}$, takes, in the proposed scenario, the following form:
\begin{equation}\nonumber
d_l(\bar{x}) = \frac{e^{\bar{x}}}{\bar{H}_0} \int_0^{\bar{x}} 
\frac{\sqrt{\phi (x)}dx}{\left( \Omega_{m0}e^{5x/2} +  
\Omega _{\Lambda0}e^{-(1/2 - w_0x) + w_1x^2}\right)^{1/2}}
\, .%\label{lumdis}
\end{equation}
Using this luminosity distance to analyze data sets, should reliably remove the apparent variation of H0 with the red-shift.},
which determines the effective profile of $H_{0,\text{eff}}(x)$, can be numerically evaluated using (\ref{rt15}) starting from $\phi(0)=1$ and implementing the following constraints by fixing the free parameters $\lambda$, $w_0$ and $w_1$. First of all, we specify that we set $\Omega_{m0}=0.403$, $\bar{H}_{0}=73.3$ according to the analysis in \cite{Brout:2022vxf} referred to the flat $w_0w_a$CDM model. This choice is justified by the fact that the evolution of the dark energy contribution is associated with an equation of state of a $w_0w_a$ model for $z \lesssim 0.2$, as reported in Fig.\ref{rho e omega}. Furthermore, $H_{\text{CMB}}=67.4$, where $H_{\text{CMB}}=H_0(x_{\text{CMB}})$ 
corresponds to the value of the Hubble constant obtained by the Planck satellite. From the condition $H_{0,\text{eff}}(x_{\text{CMB}})=H_{\text{CMB}}$, it is possible to evaluate the parameter $\lambda$ for every pair of $(w_0, w_1)$ by inverting the relation (\ref{heffphi}). The values of $(w_0, w_1)$ are selected by ensuring the positiveness of the mass of the scalar mode associated to the value of $\phi$ close to $x=0$, which, for a curved background (we also consider the adiabatic limit where time derivatives are neglected, see \cite{Olmo:2005hc,PhysRevLett.95.261102,PhysRevD.100.084014}), takes the form
\begin{equation}
    m_{\text{eff}}^2= \frac{1}{3}\left(\phi \, \frac{d^2V}{d\phi^2}-\frac{dV}{d\phi}\right).
\end{equation}
Implementing the considerations above, we find the following values for the parameters:
\begin{equation}
    w_0=0.830\,,\qquad w_1=-0.683\,,
\end{equation}
with $\lambda=-0.00435$.

\begin{figure}
\centering
\includegraphics[width=8.6cm]{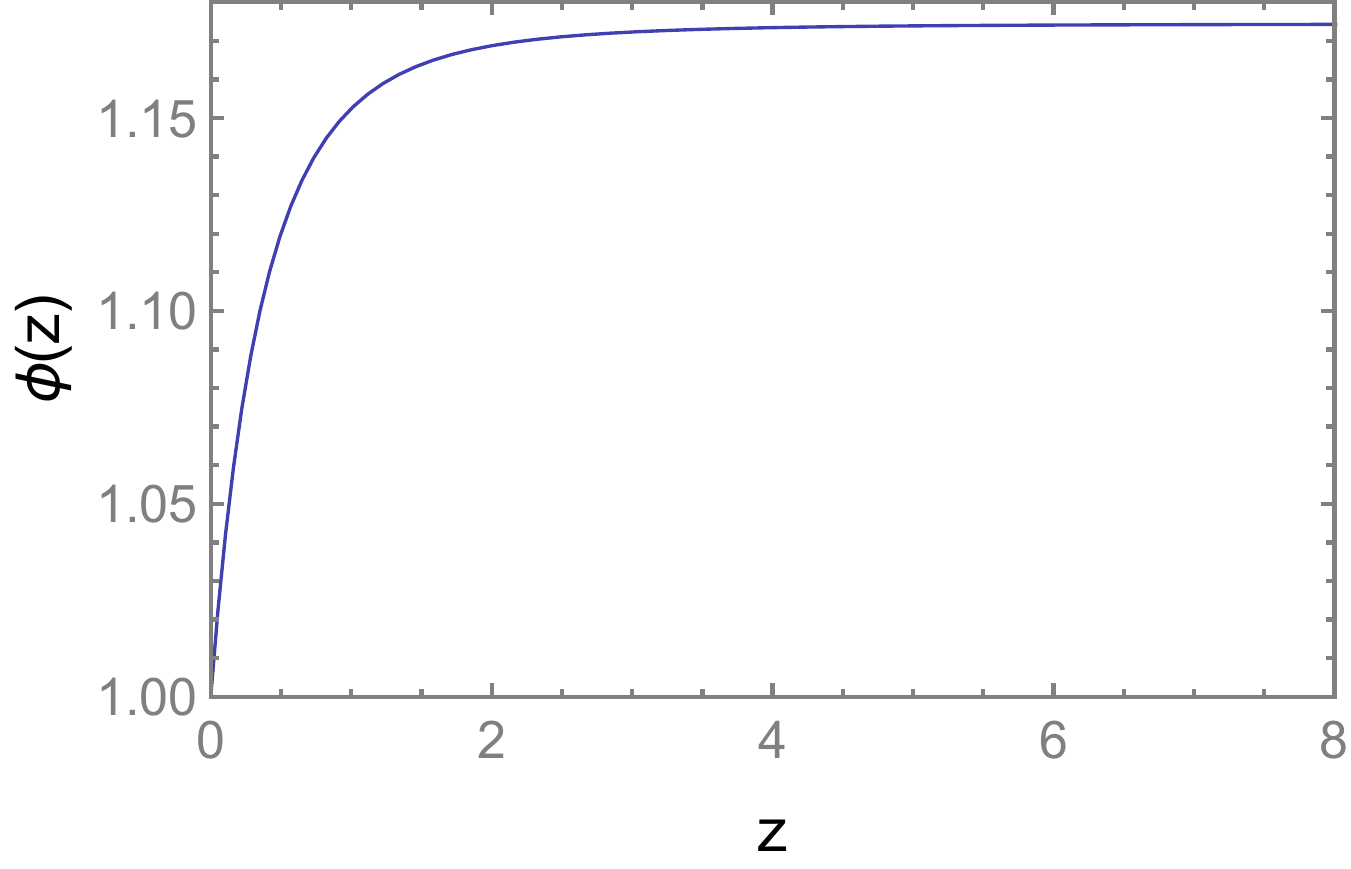}
\caption{Plot of scalar field vs. the red-shift $z$.}
\label{phi}
\end{figure}

\begin{figure}
\centering
\includegraphics[width=8.3cm]{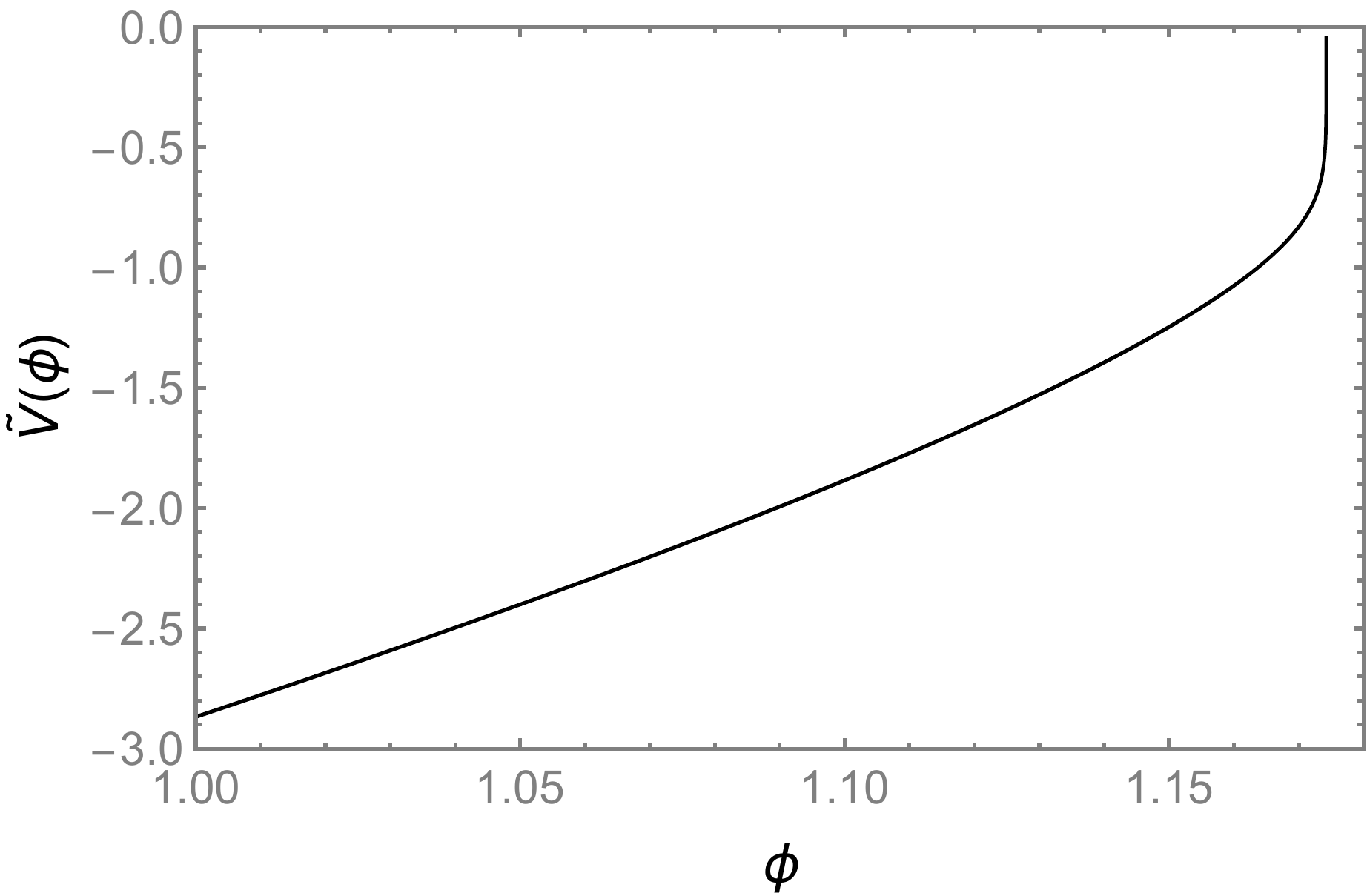}
\caption{Plot of the dimensionless potential $\tilde{V}(\phi)\equiv V(\phi)/6\bar{H}_{\text{0}}^2$. We recall that $\phi=1$ corresponds to $z=0$.}
\label{V}
\end{figure}

\begin{figure}
\centering
\includegraphics[width=8.5cm]{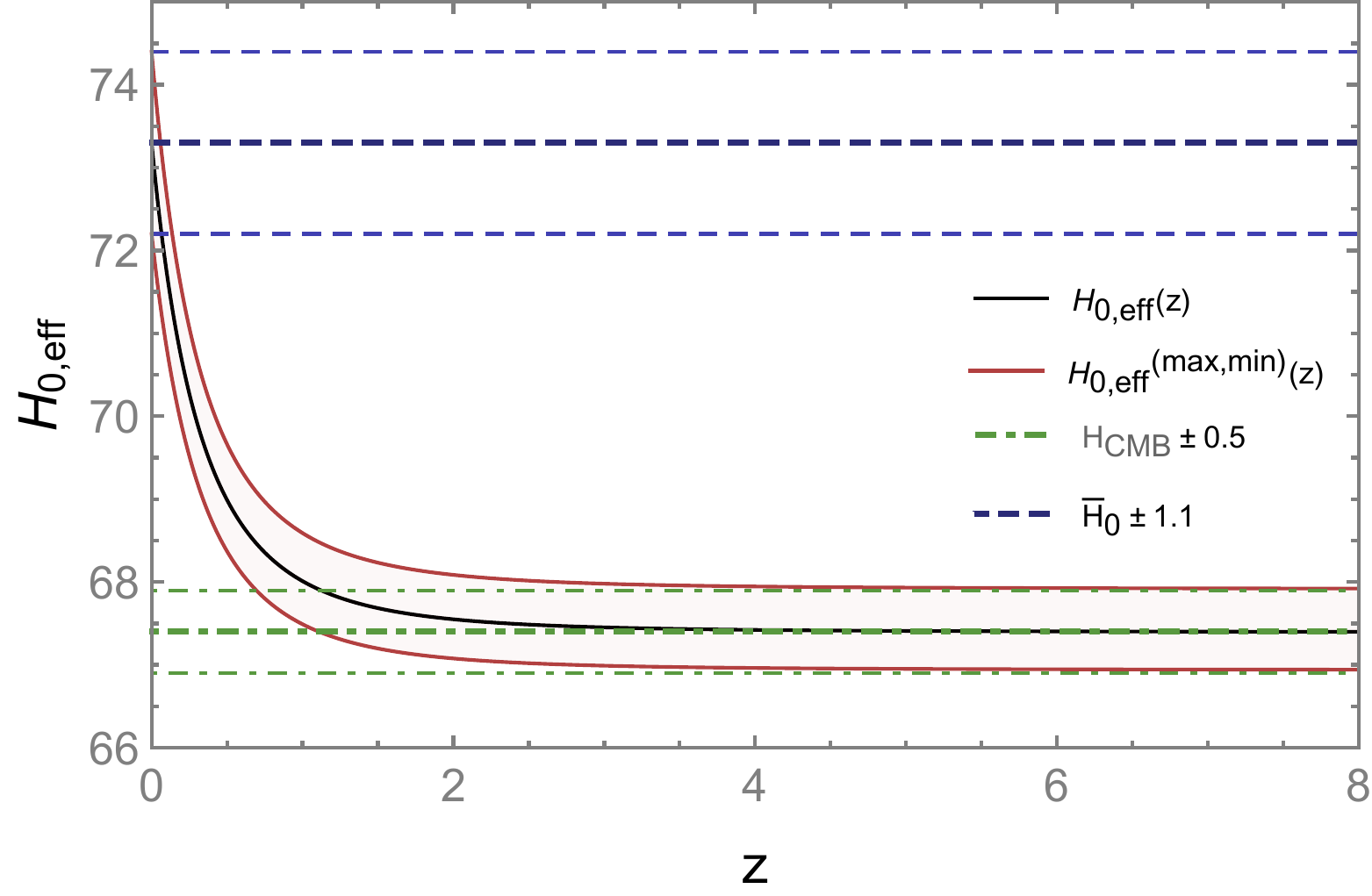}
\caption{Plot of the effective Hubble constant $H_{\text{0,eff}}$ vs. $z$.}
\label{H0ultimo}
\end{figure}

\begin{figure}
\centering
\includegraphics[width=9cm]{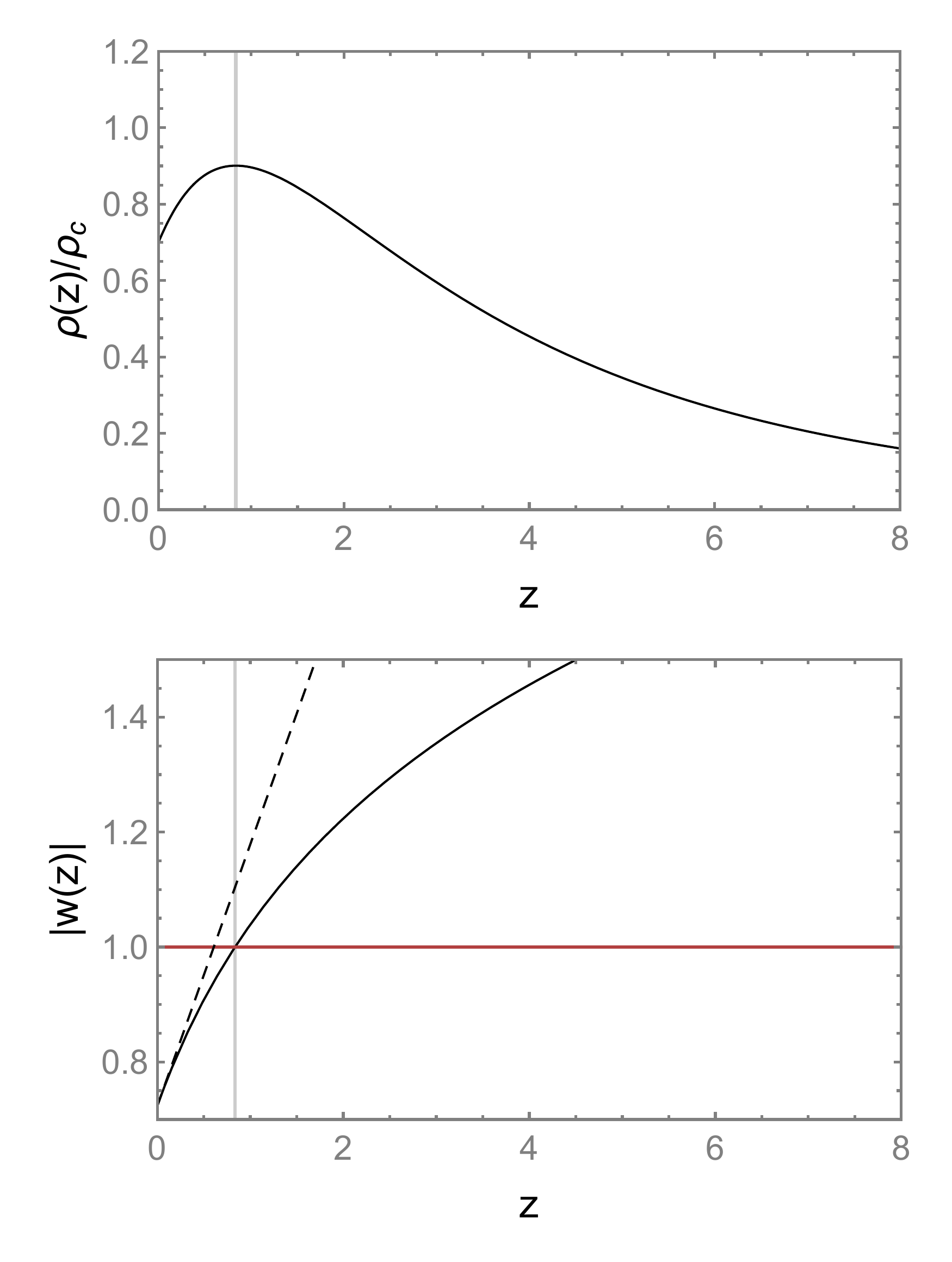}
\caption{Plot of the normalized dynamical dark energy density vs. $z$, compared with the evolution of the (modulus of) effective $w(z)$ parameter. The vertical grey line denotes the $z$ value where the transition from quintessence to phantom configuration takes place, depicted by the red line at $|w|=1$. The dashed line represents the behaviour in the limit $\log (1+z) \simeq z$, and it can be appreciated as the two curves diverge significantly starting from $z\simeq 0.2$, where the $w_0 w_a$ description does not hold anymore (see discussion in Sec.~\ref{sec: sec. 3}).}
\label{rho e omega}
\end{figure}

To clarify the achieved outcomes, let us now move to the standard red-shift variable $z$. The behaviour of $\phi(z)$ can be appreciated in Fig.~\ref{phi}. It can be noticed how General Relativity is rapidly recovered for $z\gtrsim 2$, where the scalar field boils down to a constant value. 
The potential term $V(\phi)$ is instead plotted in Fig.~\ref{V}, and it is obtained from \eqref{rt14} by numerically inverting the scalar field in $x=x(\phi)$. A few comments are now in order. The behaviour of $V(\phi)$ displayed by Fig.~\ref{V} shows a significant resemblance with the results presented in \cite{schiavone_mnras}, where a concave up profile for $V(\phi)$ is likewise obtained. In both cases, in particular, the saturation effect exhibited by $\phi$ with the redshift leads to a rapid growth of the potential for values of $\phi$ slightly larger than unity. Near $\phi\simeq 1$, however, the requirement in \cite{schiavone_mnras} of reproducing the $\Lambda$CDM model and consisting in a nearly flat potential results in a depart from Fig.~\ref{V}, which is instead related to a $w_0 w_a$CDM scenario. A flat region for $\phi\simeq 1$ is also present in Fig.~3 of \cite{Dainotti:2022bzg}, which is in addition endowed with a concave down-trend, leading to a decreasing of the potential for $\phi>1$. 
\newline Equation (\ref{heffphi}) finally provides the obtained behaviour of $H_{0,\text{eff}}(z)$ which corresponds to the central profile depicted in Fig.~\ref{H0ultimo}.

In Fig.~\ref{H0ultimo}, we plot also the maximal and minimal curves of $H_{\text{0,eff}}$, which enclose the red-shaded region and are obtained by considering the errors of Planck and SNIa measurements: $\bar{H}_{0}=73.3\pm1.1$ and $H_{\text{CMB}}=67.4\pm0.5$. In fact, using the same approach described above, we find the following values for $(w_0,w_1)$: 
\begin{align}
w_{0,\text{max}}=0.905,\qquad w_{1,\text{max}}=-0.860,\\
w_{0,\text{min}}=0.651,\qquad w_{1,\text{min}}=-0.627,
\end{align}
with $\lambda_{\text{max}}=-0.00393$ and $\lambda_{\text{min}}=-0.0049$. We want to stress that the same analysis can be performed by using different values of the cosmological parameters $\Omega_{m0}$ and $\bar{H}_0$ yielding very similar quantitative results. A  detailed comparison of cosmological constraints from combinations of probes and different cosmological models can be found in \cite{Pan-STARRS1:2017jku,Brout:2022vxf}.

As a result, the non-minimally coupled scalar field is actually responsible for the scaling of the effective Hubble constant, which mostly occurs for $z\lesssim 2$, where the quintessence properties of the dynamical dark energy fluid (see Fig.~\ref{rho e omega}) ensure that no tachyon modes emerge in the theory. At higher red-shift, the underlying gravitational scenario approaches General Relativity (up to a rescaling of the gravitational constant given by the asymptotic value of the scalar field, see Fig.~\ref{phi}), and a transition from quintessence to phantom dark energy model ($w<-1$) takes place at $z\simeq0.83$, where $\rho(z)$ reaches its maximum.

%\subsection{The f(R) profile at low red-shift}\label{subsec: subsec. 4}
To conclude our analysis, we present an approximated solution for the $f(R)$ function under the low-red-shift condition ($z\lesssim 1$). To recover the explicit behaviour of the function $f(R)$ defined in \eqref{rt2}, according to the analysis above, %we proceed with a complete numerical analysis of the system \eqref{rt10}-\eqref{rt12}-\eqref{rt11}, which we solve in terms of $\phi(x)$ and $V(x)$. Thus, 
we perform the inversion $x=x(\phi)$ enabling us to obtain $R = dV/d\phi$ and $\phi = \phi(R)$. Finally, by fitting such a numerical solution up to $z=0.5$ and considering a profile of the form
\begin{equation}
    f(R)=c_1 + R + c_2 R^2,
\end{equation}
where constant, linear and quadratic terms in R are included. The recovery of the $\Lambda$CDM model occurs when $\phi = const.$ (defined as $\phi=df/dR$), thus obtaining the values $c_1=2.72\, \bar{H}_0^2$ and $c_2=0.0045\,\bar{H}_0^{-2}$. As a check for the theory, the positive $c_2$ coefficient in front of the $R^2$ term ensures the absence of a tachyonic field and well-behaved cosmological solutions with a proper era of matter domination (\cite{Sawicki:2007tf}).

\section{Conclusions}\label{sec: sec. 6}
In this work, we assumed that the Hubble tension is a consolidated feature, emerging from the comparison of SNIa Pantheon and Pantheon+ data (\cite{Scolnic_2022,Pan-STARRS1:2017jku,Huang_2020,Scolnic:2023mrv,Murakami:2023xuy} with CMB data \cite{Planck:2018vyg}) (see also \cite{Dainotti:2023ebr} for a reduction of the error bars in SNe Ia). As argued in \cite{Dainotti:2021pqg,galaxies10010024}, the Hubble tension can be associated to an effective $H_0(z)$ profile, as it emerges from the SNIa sample and is reliable up to the recombination red-shift $z\simeq 1100$ (for a theoretical representation of this scenario via a modified gravity theory, see also \cite{schiavone_mnras}). The present study provides an effective $H_0(z)$ behaviour that rapidly decays across the SNIa samples, approaching Planck measurements via a plateau for $z\sim 5$. This scenario has been reproduced by combining a metric $f(R)$ gravity, which controls the dynamics at low red-shift, together with a dynamical dark energy density exhibiting a phantom divide line for $z\gtrsim 0.83$. The whole of these effects concurs in driving the desired dynamical picture, with the non-minimally coupled scalar field responsible for the rescaling of the Hubble constant, and the dynamical dark energy source guaranteeing the absence of tachyonic instabilities, otherwise affecting the considered modified gravity model at low red-shifts. The proposed scenario has the merit of providing a precise marker for its validation, since it discriminates low red-shift sources, like SNIa, from larger ones, like QUASARS (\cite{Colgain:2022nlb,Dainotti:2022rfz,Bargiacchi:2023jse}) and Gamma Ray Bursts (\cite{Dainotti:2022wli,10.1093/pasj/psac057,Cao:2022wlg,Cao:2022yvi,10.1093/mnras/stac2752}). Within the error bars, indeed, there is a possible agreement between our predictions and data sets from distant sources, see e.g. \cite{Dainotti_2020}, bearing in mind that the increase of the statistics in the samples of high $z$ sources will allow more accurate comparison with our model. At this stage, we simply observe that by taking into account error bars of $H_0$ measurements, the behaviour in Fig.~\ref{H0ultimo} is consistent with Planck data already for $z\simeq 2$, suggesting that the $H_0(z)$ profile could mainly run within the red-shift interval available to SNIa samples.

\section*{acknowledgment}
The work of FB is supported by the postdoctoral grant CIAPOS/2021/169. MDA is supported by an EPSRC studentship. Authors thank Gonzalo J. Olmo for the useful comments and Maria Giovanna Dainotti for valuable suggestions on Fig.\ref{H0ultimo}.

\section*{Data Availability}
No new data were generated or analysed in support of this research.

\bibliographystyle{mnras}
\bibliography{main}

\bsp	% typesetting comment
\label{lastpage}
\end{document}